\documentclass[10pt]{article}
\usepackage{graphicx}

\usepackage{latexsym}
\usepackage{amsmath}
\begin{document}

\title{Gompertz and Verhulst frameworks for growth AND decay description}
\author{Marcel  AUSLOOS  \\ 
 \\ {\it now at} R\' esidence Beauvallon, 
  rue de la Belle Jardini\`ere, \\B-4031 Angleur-Li\`ege, $ still$ Belgium \\ 
fax : (int 32) 4 371 43 40\\
$e$-$mail$ $address$:
marcel.ausloos@ulg.ac.be}

\begin{abstract}
Verhulst logistic curve either grows OR decays, depending on the {\it growth rate} parameter value. A similar situation is found in the Gompertz law about human mortality. However, growth does not have to be infinite nor reach a  finite steady state at an infinite asymptotic time.  Moreover, before some decay, some growth must have occurred.  For example, some leveling-off could occur at finite time,  followed either by some growth again or then by some decay. Numerous examples show that the Verhulst and Gompertz models are   too reductive (or restrictive) descriptions of their original purpose.   It is aimed, in the present note, to encompass into ONE simple differential equation the growth AND decay features of,  e.g., population sizes, or numbers, but also of many  other measured characteristics found in social and physical science systems. Previous generalizations of Verhulst or Gompertz functions are recalled. It is shown that drastic growth or decay jumps or turnovers can be  readily described through drastic changes in values of the growth or decay rate.  However smoother descriptions can be found if  the growth or decay rate is modified in order to take into account some time or size dependence. Similar arguments can be carried through, but not so easily,  for the so called carrying capacity, indeed leading to more elaborate algebraic work.\\ 

\end{abstract}
 
 \maketitle

\section{Introduction}

 Verhulst  1845  seminal paper \cite{Verhulst845} bore on previous  (1798) considerations by Malthus \cite{Malthus798} on population $size$ evolution.
 Verhulst model attempts to describe how the.growth (or {\it mutatis mutandis} its decay) of a $single$ species towards its equilibrium population is limited, taking into account the reproduction rate $r$ and the nowadays so called {\it limiting capacity} $M$, i.e.,  the maximum value the   population can reach in its  environment., due to limited resources. He obtained what he called the logistic curve (or map). In a subsequent paper, Verhulst \cite{Verhulst847} attempted a  different mathematical form of the blocking growth term, thereby obtaining an exponentially slow approach of the carrying capacity, but not   a sigmoid.
 
There have been many considerations  starting from Verhulst approach of population $size$ evolution, generalising his original work. However, it is of interest to recall that 
among the generalisations, one should distinguish between those which stick to one single differential equation, and those which involve a set of equations, because they consider a competition effect with another population.   Among the former ones,  I am fond of the  1939 work by Avrami  \cite{Avrami39} on crystal growth, - the so called species changing its thermodynamic phase or state, but there are many others. Among the latter, one has only to  go back to the work of  
Lotka, published   \cite{Lotka25}, in 1925, and of  Volterra \cite{Volterra31}, in 1931, pertaining to so called {\it prey-predator models}. Not attempting to cite here  many  subsequent publications., let it be pointed out that Blanco  \cite{Blanco1}  has suggested a trivial  relationship between the logistic equation and the Lotka-Volterra (LV) models,   indicating  that the former is in fact a sort of limit of the LV equations, pending a redefinition of the parameters.

However, growth can neither be infinite nor reach a  finite steady state at an infinite asymptotic time, as in the logistic map.  Moreover before some decay, some growth must have occurred. In particular, some leveling-off could occur at finite time,   though followed either by some growth again or then by some decay, in both  cases, with a smooth or an oscillatory variation. 

It is of interest in contrast, to recall  ideas byGompertz   \cite{Gompertz825}  who in  a series of papers near 1825 discussed  'the Nature of the Function Expressive of the Law of Human Mortality, and (...) a New Mode of Determining the Value of Life Contingencies''. 
The   Gompertz model for describing the decreasing number of members of a population can be in fact easily reformulated in terms of a (simple first order differential)  $growth$ equation of Malthusian form, with an {\it exponentially decaying birth rate}.  This leads to another form of an analytical sigmoid curve,  though asymmetric,  in contrast to the Verhulst logistic map.

\begin{figure}
\includegraphics
[scale=0.4]{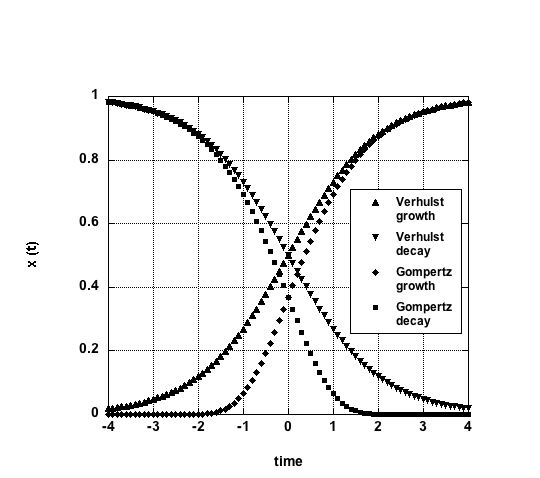}
\includegraphics 
[scale=0.4]{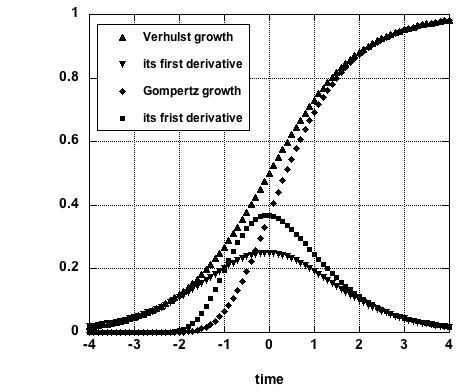}
\caption{\label{V1G1} Comparing a Verhulst and a Gompertz  growth or decay law (lhs), and their first derivative, but only in the case of parameter $r>0$ for growth in  (rhs); see the asymmetry in Gompertz laws}
\end{figure}

In brief, depending on the sign of the parameters in Verhulst or Gompertz approach, one can describe some growth or some decay see Fig. \ref{V1G1}  (lhs). However numerous examples show that  both  Verhulst and Gompertz modelisations are thus   too reductive (or restrictive) descriptions of their original purpose.  Whence, it is of interest to aim at  finding some simple description combining some less trivial behaviour,  i.e. obviously, practically inducing some extremum at some time in the function which should be  the solution of a, by request, simple first order differential equation.   This is the aim of this note, i.e.
to encompass into ONE simple first order differential equation {\it  growth AND decay} features of,  e.g., population sizes, or numbers, but also of many  other measured characteristics found in social and physical science systems.

In Sect. 2,  after recalling Malthus and Verhulst equations, a few applications are mentioned in order  to substantiate the broad framework in which Verhulst equation can be applied.  $Growth$, indeed, is not only found in population studies, through counting the number of individuals in some system, but can be $measured$ through other means, like the number of stems, the mass, the length, ... of ''something''. It is often found, of course, that the Verhulst sigmoid is only a rough representation of the data. Whence one has sometimes invented artificially skewed logistic curves  \cite{PearlReed23}; see Sect. 2, for a general introduction to $growth$ and such subsequent  considerations.
 
 Subsequent  considerations on decay, with examples about languages and religions, are to be found in Sect. 3, before outlining Gompertz law.  Discussion about  bio-demography applications can be found in an interesting  1997 review, up by Olshansky and Carnes  \cite{OlshanskyCarnes}.
 Nevertheless, {\it ad hoc}, previous generalisations of Verhulst or Gompertz functions are briefly recalled
to conclude Sect. 3.
  
In Sect. 4, a few examples show that the Verhulst and Gompertz modelizations are  indeed too reductive (or restrictive) descriptions of reality. However, 
drastic growth or decay jumps or evolution turnovers can be  readily described through drastic changes in values of the growth or decay rate, as shown inSect. 5.  Sometimes the evolution show drastic jumps or drops indeed, - due to exogenous or endogenous causes. However smoother descriptions must be found, i.e. combining some less trivial behaviour.  Obviously, one should aim at practically inducing some extremum at some time in a smooth function which should be  the solution of a, by request, simple  first order differential equation.  

It is aimed, in Sect. 6, to encompass into such ONE simple differential equation the growth AND decay features of population sizes, or numbers, but also of many  other measured characteristics found in social and physical science systems. 
In fact one can observe that one could consider that he growth or decay rate has  either some time or some size dependence.  The algebra is rather simple in order to recover a Verhulst or Gompertz evolution equation..
In Sect. 7,   the time and/or size dependence of the so called carrying capacity is considered in the same spirit as the growth rate. However this is less trivial, indeed leading to more elaborate algebraic work.
  Sect. 8 serves as a conclusion.
  
\section{Introduction to Growth}

Following the above  historical points, let us do some mathematics and data  analysis within Verhulst  equation framework.

\subsection{Growth equations}

Recall Malthus model or rather his equation for  describing a (normalized) population  $size$  $x$ evolution as a function of time $t$,  due to a birth rate $r$, i.e. 
\begin{equation} \label {Malthus1diff}
 \frac{dx}{dt}= r \; x,   
\end{equation}
which has the (catastrophic) exponential solution
$x(t) =e^{rt}.$

The original Verhulst  population growth model considers that resources are limited, such that the evolution of the number of members $N(t)$ in the population is described  in terms of a carrying capacity $M$  and a growth rate $r$,  according to
\begin{equation} \label{Verhulsteq19}
\frac{dN}{dt}=\;r \; N \;  \left[ 1 -       \frac{N}{M} \right]\;.
\end{equation}

With the change of variable $x= N/M$, if  $M$ is a constant,     
 a  dimensionless form equation can be thereby written, i.e.
\begin{equation} \label {Verhulst1diff}
\frac{dx}{dt}=r  \; x \;  \left[ 1 -  x \right]\;     .
\end{equation}
It has for solution, the so called  logistic map, a sigmoid curve, 
\begin{equation} \label {Verhulst2sol}
x=\frac{e^{rt}}{1+e^{rt}}\;,
\end{equation}
if $r$ is a constant as well.
  Notice  that Verhulst had examined  complicated forms of the growth  limiting  term,    like a square, 
but also a  square root \cite{Verhulst845}, both cases   without  any theoretical justification, but aiming at better fitting available data
 Indeed, there is no biological justification for this assumption; Gilpin, Case, \& Ayala \cite{Gilipinetla66} proposed a growth equation of the form $d x/dt = r\;x [1- (x/M)^{\theta} $, - a larger than 1  value of $\theta$ resulting in a steepening of the growth curve and a higher value for the inflection point, as easily calculated.
   
\subsection{Applications of  the logistic map}

 Let us point out  to  a few cases of the  ''Verhulst population growth model'', as discussed during the oral presentation, in order to show   some universality value of the logistic map application.
 \begin{itemize}
\item Population size:
Beside those considered by Verhulst, an  interesting application  on population size is that of 
Montroll  and  Badger \cite{MontrollBadger74}, in 1974, who studied the USA population between 1810 and 1970.  An interesting parameter value is found, i.e.
$r = 0.02984$, which implies interesting considerations on  american sexuality  during those  times.
\item Mass growth:  
However the logistic map can also be applied to, e.g.,  mass growth, like in the report of  Lee  and Werning \cite{LeeWerning08}, in 2008,  who compared
Alligators,  Elephants, Shrews 	and	  Ostriches, on one hand, but in the other hands  Tenontosauruses, Allosauruses, and Tyrannosauruses, using skeletochronology; see   reference    \cite{LeeWerning08} for the parameter $r$ values,   in $kg/yr$, and subsequent discussions. For completeness, let us mention that the growth curves were compared with an alternative model to that of Verhulst, i.e.  the von Bertalanffy  \cite{vonBertalanffy}  curve  of mass accumulation.
\item Length growth: As far as 1928, Pearl group studied the 
Cucumis Melo stem length, either under field  conditions   \cite{Pearletal28f} or  in the absence of exogenous food and light
 \cite{Pearletal28a}. The $population$ was considered to be the number of nodes. Notice that $r$   is of the order of 0.7 in their observations.
 \end{itemize}

\subsection{Skewed logistic}

In so doing,  i.e., studying Canteloup growth, Pearl et al.  empirically found that one should generalised the logistic map in order to reproduce better the data. Rather than the mere logistic, they propose a form like 
\begin{eqnarray} \label{skewedlog1expans}
 y(t)=\frac{k}{1+e^{a_0+a_1\;t+a_2\;t^2+...}}      \; ,    
\end{eqnarray}
where $y$, in  \cite{Pearletal28a}, is the number of seedlings of the canteloups.
 
 It is $quasi$  obvious that a more general form is like
\begin{eqnarray} \label{skewedlog1integr}  
 y(t)=\frac{k}{1+e^{f(t)}}      
\end{eqnarray}
to be discussed below.

\section{ Introduction to Decay}

Recent applications along Verhulst  (limited growth) approach have been concerned with  the $decay$ of a population size, rather than its growth.

\subsection{Religions}
The time evolution of several ''main'' religions was considered to be described, at  a so called microscopic level,    along the lines of
the  Avrami-Kolmogorov equation, describing a liquid-solid state formation in a continuous time framework \cite{auslooscrystalgrowth}, i.e. a peculiar form of Verhulst law \cite{DeVriesetal03}. The solution   is usually written as
\begin{equation}\label{Avrami}
c(t)=1- e^{- K t^{n} }
\end{equation}
where $c(t)$ is the volume fraction being transformed from one phase to another; $K$  and $n$ are adjustable parameters.  However, it is ''obvious''  that before decaying the  number of adepts of any religion had to grow from  the founder, $N(0)=1$. Nowadays, several  ''religions''  have still  a growing number of  adepts. Others have reached a maximum, thus decay,  or go again into a growing pattern. Population growth-death equations have thus been conjectured  for modelling  such religion  $size$
evolution dynamics, in a continuous time framework \cite{hayw99}, \cite{hayw05}, \cite{religion1}, \cite{religion2}, \cite{religion566} and \cite{religion568} attempting to describe the existence of maxima in the number of adepts. An interpretation of the features, after data fitting, can resort to an attachment parameter, as  resulting from  joining at first one's  ''mother religion'', within exogenous field conditions,  as one ''joins'' at first one's mother tongue.

\subsection{Language death}

For completeness, notice that the disappearance of world's languages has also been  recently considered through a Verhulst  time  evolution   equation in terms of the number density $\rho_i$ of practitioners of some language $i$, in \cite{NowakKrakauerPNAS99} and \cite{Abr+03},  i.e.,
\begin{equation}\label{verhulst}
\frac{\partial \rho_i}{\partial t} = r_i\; \rho_i \left( 1- \frac{\rho_i}{C_i} \right)
\end{equation}
where $C_i$
is the carrying capacity of the environment  for the population speaking the $i$ language
and $r_i$
is a  ''negative growth'' rate.  The $r_i$ values are at this level of no interest. However notice that the  speakers do not fully represent   a single so called population, since their member evolution is toward   a bilingual attitude. In fact, there is a $competition$ between two (or more) languages/populations, - which according to the point of view  mentioned in  the Introduction   rather pertains to  a prey-predator problem.

\section{Gompertz Death  law}

In fact, previous to Verhulst, Gompertz  \cite{Gompertz825}  discussed human  population mortality dynamics  from a practicing actuary point of view.  Gompertz used equal long time intervals, i.e. 10 year long intervals,  to observe that the differences between the common logarithm of the number of persons living in  such successive equal age intervals were almost identical during a significant portion of their life span. Thus  the numbers of living individuals in each yearly increase of age are   in geometrical progression. Whence mortality progresses geometrically as age increases arithmetically.  Therefore Gompertz proposed a 3 parameter expression for the number $N(t)$ of $survivors$ of a population
at time $ t$:
\begin{equation} \label {Gompertz2sol}
N(t) / N_{inf} = \; exp [-b \;exp (-rt)],
\end{equation}
where $N_{inf}$,  $b$ and $r$ are positive constants.  This corresponds to  an exponentially decaying birth rate $r$ in Malthus equation, i.e.
$ r= r_0 e^{-\kappa t}$, pending $r_0$ and $\kappa$ being positive constants. For completeness, notice that Makeham introduced an age independent mortality rate to be considered in order to better represent available data \cite{Makeham},  i.e. adding a $A+Bt$ term in the right hand side of Eq.(\ref{Gompertz2sol}, thus attempting  to take into account extrinsic effects.

The easily derived differential equation describing the exponential rise in death rates between sexual maturity and old age, is commonly referred to as the Gompertz equation,  i.e., 
\begin{equation} \label {Gompertz1dif}
\frac{dx}{dt}= r \; x \;  log\left[  \frac{k}{x} \right]\;\\
\end{equation}
where $k$ has {\it mutatis mutandis}  the same meaning as $M$, the carrying capacity, in the Verhulst approach. Notice that  Gompertz and Makeham wondered whether   $k$  can be interpreted in a socio-biological framework,  i.e., as  a biological limit to life, based on some {\it vital force}.

\subsection{Skewed Gompertz law}

 In the original Gompertz model, $r$ and $k$ are supposed to be constant;  $b$, in Eq.(\ref{Gompertz2sol}), and $k$, in Eq.(\ref{Gompertz1dif}), are related to each other through $r$ and the initial (or final) conditions on $N(t)$. In 1924,  Trachtenberg \cite{Trachtenberg} already  examined 
 extensions of Gompertz law, sometimes departing from Gompertz   geometric progression hypothesis, - extensions  which do not necessarily lead to a sigmoid curve,  but   can be called skewed Gompertz laws.
 One of them reads
 \begin{equation}\label{Trachtenberg}
N(t) = N_{inf} \; e^{ [a+bt +ct^2+dt^3+et^4]}\;. 
 \end{equation}  It can  clearly be the solution of a second order differential equation for $log(N(t))$, as wished by Gompertz, and   presents two inflexion points at precise $t_{\alpha}$ and  $t_{\beta}$.
 
 In fact, similarly to  Eq.(\ref{skewedlog1integr}),  the most general  extension of a Gompertz form can be hereby proposed to be
\begin{eqnarray} \label{skewedG1integr}
N(t) = N_{inf} \; exp [-b \;exp (-g(t))]\; .
\end{eqnarray}
  
 \subsection{Biomedical considerations}
 
 For the purpose of the reader general information,
 an interesting  1997 review by Olshansky and Carnes  \cite{OlshanskyCarnes},  on ''ever since Gompertz'', should be quoted. However, since the 
 discussion  is mainly about   demography applications  and biological causes of mortality, in various species,  - like the vital force  concept and in particular the influence of radiations, the paper content is quite aside our present considerations. and not further elaborated upon, though of great interest.

\begin{figure}
\includegraphics[height=8cm,width=18cm]{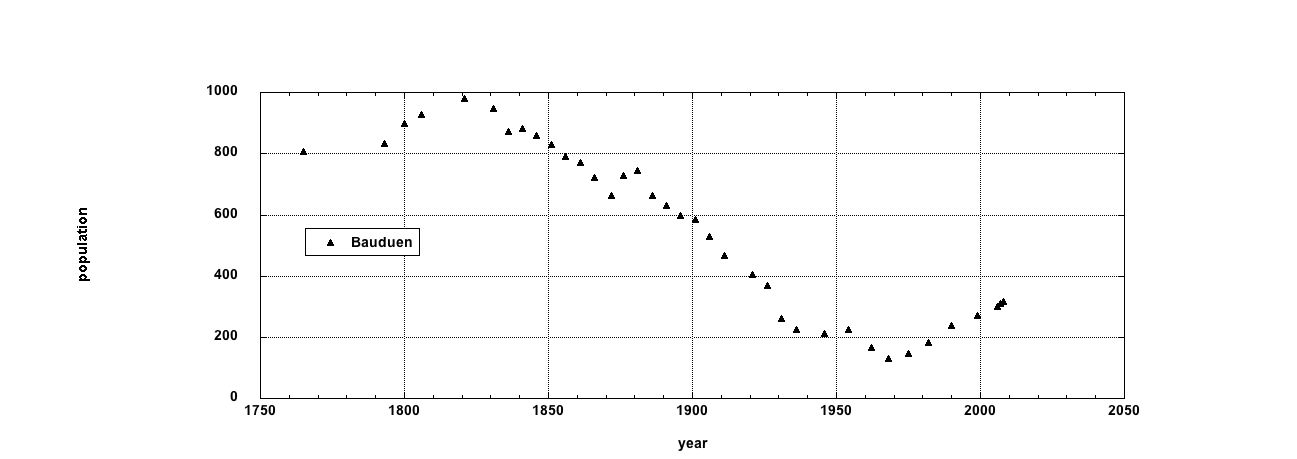} 
\caption{\label{bauduen} Evolution of the population in Bauduen, Var since $ca.$ 1750}
\end{figure}

Nevertheless, it is of common knowledge that $exogenous$ causes can  manifest themselves and influence some "decay" in the number of living individuals  in a population. Take the trout population data of Lake Ontario measured by Beeton 
\cite{Beeton70}, quoted by Meadows et al. \cite{Meadows}, for example. 
Pollution is supposed to be the influential external field. 
  On the other hand, a case of $endogenous$-like cause was noticed by Kormandy \cite{Kormandy69}, in 1969,  who commented on data, which I did not find,  and who  stated that   deer or goats,  even though natural enemies  are absent, "often overgraze their range  and cause erosion  or destruction  of the vegetation'',  and consequently  die. 
Finally, on the  Bauduen, Var  city wikipedia website  ($http://fr.wikipedia.org/wiki/Bauduen$) it is reported that the city population, since 1793,  increased till 1820, thereafter decreased till 1970 and is going back up nowadays. In this case, sharp turnovers are found at the time of war ends; see Fig. \ref{bauduen}. 

\section{Growth and decay and growth and decay and ...}

  In the case of Lake Ontario trouts, here above mentioned \cite{Beeton70}, there is obviously a maximum near 1925, followed by a rather sharp drop of three orders of magnitude in 1955 or so.  Similar though smoother,  features are found for Bauduen population (Fig. \ref{bauduen}). In brief, such few examples should remind us that, in many measurements, when counting  the evolution of a population (of whatever),  there is a  succession of maxima and minima,  though the growth and decay regimes can be of different sizes and time spans. Remember business cycles as well \cite{ormerodPhA293}, \cite{miskiewiczauslooscycles}, \cite{sanglierauslooscycles},  \cite{Wright05},  or \cite{redelicoausloos08}. To debate on the evolution laws $per$ $se$ is out of question here, but to find an equation containing both  types of extrema is the goal of this report, basing our considerations on either Verhulst  or Gompertz equation.

\subsection{Beyond Verhulst and Gompertz equations}

In view of the above, it is obvious that many multiparameter extensions  have been proposed in order
 to ''improve'' Verhulst or Gompertz laws  \cite{AhujaNash67}, see App. A. However,   parameter justification is often missing, within our present requirements.  Thus a simple mathematical framework to find several inflexion points on a time dependent ''growth'' curve is still of interest.  One may imagine to introduce extra terms, representing endogenous or/and exogenous causes  for growth or decay  in Verhulst or Gompertz formalisms, Let us mention Bass model \cite{Bass} of innovative and imitative behaviour of consumers,  or Zwanzig introduction of a (damped exponential) heredity kernel  \cite{ZwanzigPNAS70}, both generalising Verhulst approach.   There is to my knowledge no equivalent extension within Gompertz framework. However both, Bass and Zwanzig considerations, have more physical bases than other purely mathematically oriented extensions.

For completeness, let us recall Bass model \cite{Bass}   of new product acquisition by consumers; it leads to a  single ''diffusion''   
equation with constant coefficients, - perfectly suitable to our aims:
 \begin{equation} \label{Bassdiff}
\frac{dX}{dt} = [N -X(t)]\;[p + (\frac{q}{N}) X(t)]\;.
 \end{equation} 
where $X(t)$ is the number of product users at time $t$; $N$ is the number of
potential users, like the carrying capacity previously, $p$ is the so called coefficient of innovation, and $q$ is  the so called  coefficient of
imitation. It is a Verhulst-like equation, though the r.h.s has an extra zero order term.
The solution reads
\begin{equation}\label{Bassol}
 X(t)= N \frac{1-e^{-(p+q)t}}{1+ (q/p)e^{-(p+q)t} }\;.
\end{equation} By differentiation, one can  easily find the time at which the function has a maximum and its value.

On the other hand, Zwanzig \cite{ZwanzigPNAS70}   writing of Verhulst equation amounts to a change of variable $e^v=N/M$,  - somewhat as in Montroll \cite{MontrollPNAS69} considering that the evolution concerns $log(N)$ rather than $N$,  and the introduction of a heredity kernel $K(t)=K(0)\;e^{t/T}$, such that one obtains an integro-differential equation (not written here). $K(0)T$ is in fact like the growth rate $r$ in Eq.(\ref{Verhulsteq19}).  For finite $T$, one obtains an oscillatory but damped evolution, with an overshoot of the carrying capacity value in fact; see Fig. 1 in \cite{ZwanzigPNAS70}.

Both previous formalisms  present at least one  hump in the ''population'' evolution. Nevertheless already within  e.g. Verhulst framework,  it can be observed that describing either (i) sharp turn-overs or (ii) drop or jump transitions is possible. It is for example, simple to introduce a drastic change in the $r$ value.  Similar situations can be illustrated with Gompertz framework, as shown  for both cases here below.

\subsection{Sharp turn-overs}

\begin{figure}
\includegraphics[height=8cm,width=8cm]{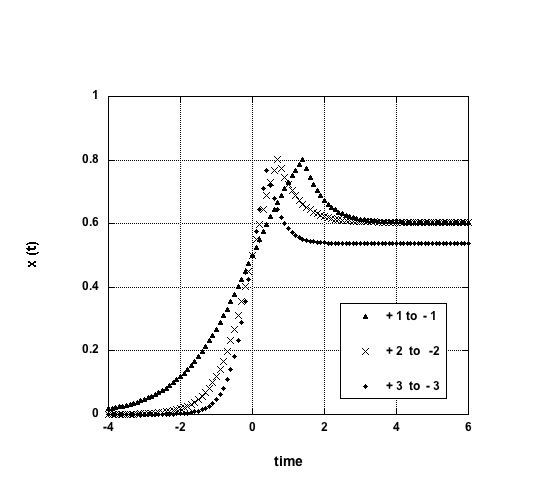}
\includegraphics[height=8cm,width=8cm]{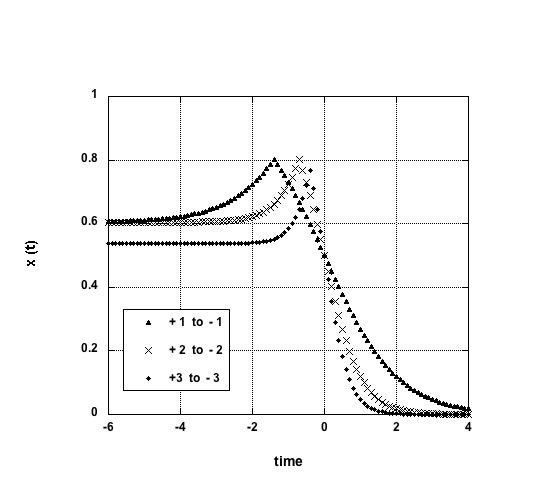}
\includegraphics[height=8cm,width=8cm]{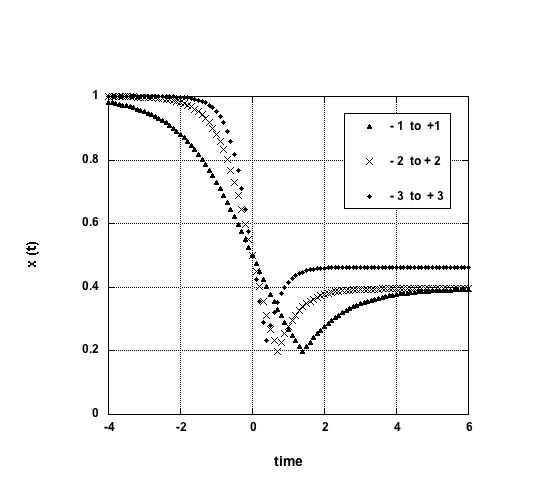}
\includegraphics[height=8cm,width=8cm]{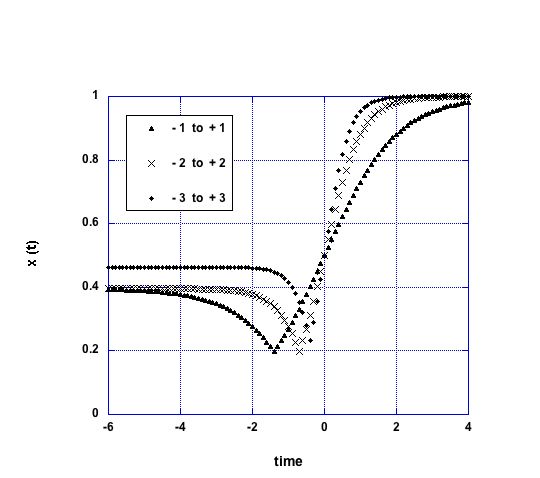}
\caption{\label{sharpturnsV} The four possible cases of sharp turn overs with continuous evolution in a Verhulst approach due to a drastic shift in $r$ values, see insert,   at different times $t$}
\end{figure}

\begin{figure}
\includegraphics[height=8cm,width=8cm]{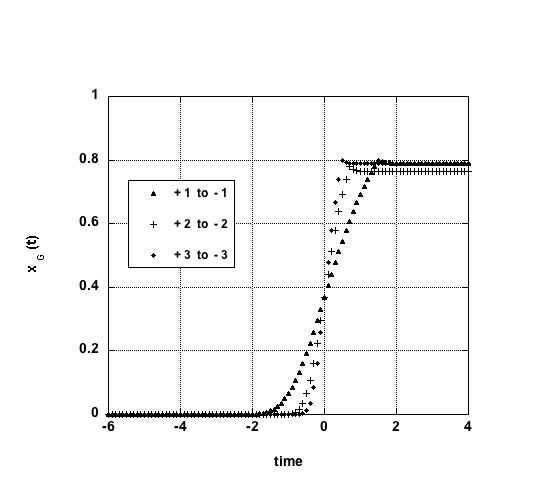}
\includegraphics[height=8cm,width=8cm]{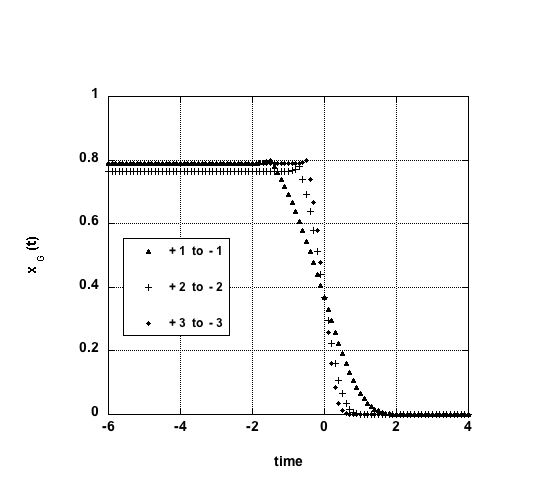}
\includegraphics[height=8cm,width=8cm]{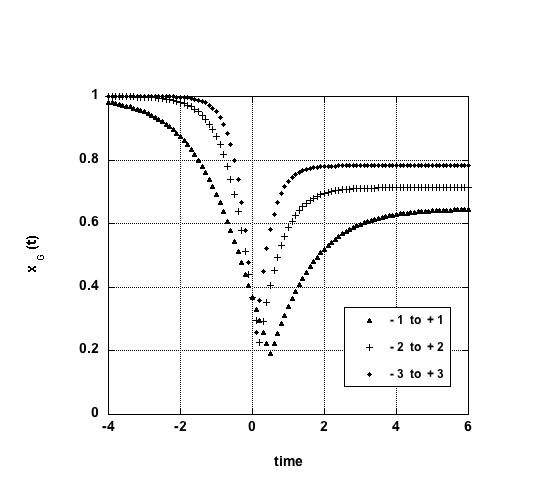}
\includegraphics[height=8cm,width=8cm]{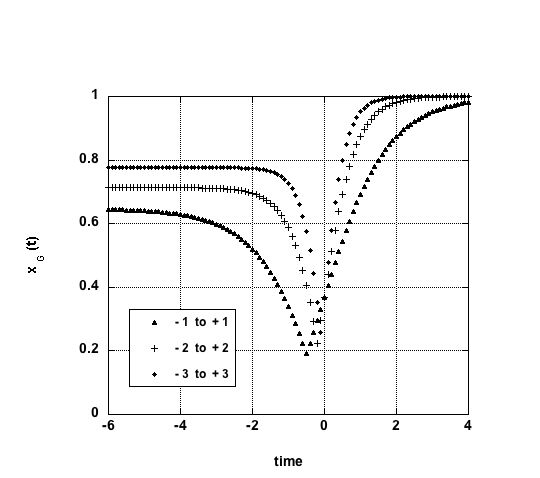}
\caption{\label{sharpturnsG} The four possible cases of sharp turn overs with continuous evolution in a Gompertz approach due to a drastic shift in $r$ values, see insert,   at various times $t$}
\end{figure}

 The four possible cases of sharp turn-overs  forcing a continuous evolution of the population number, in a Verhulst approach, due to a drastic shift in $r$ values, i.e.  from some $r$ to some $-r$ value, or the opposite, at some time $t$ are shown in Fig. \ref{sharpturnsV}. Similar curves are shown for the Gompertz evolution law on Fig.  \ref{sharpturnsG}. Notice the pronounced asymmetry in the latter case.
 
\subsection{Drop or jump transitions}

\begin{figure} 
\includegraphics[height=8cm,width=8cm]{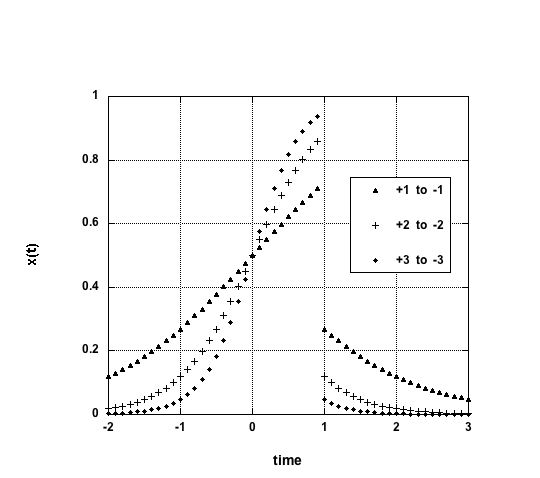}
\includegraphics[height=8cm,width=8cm]{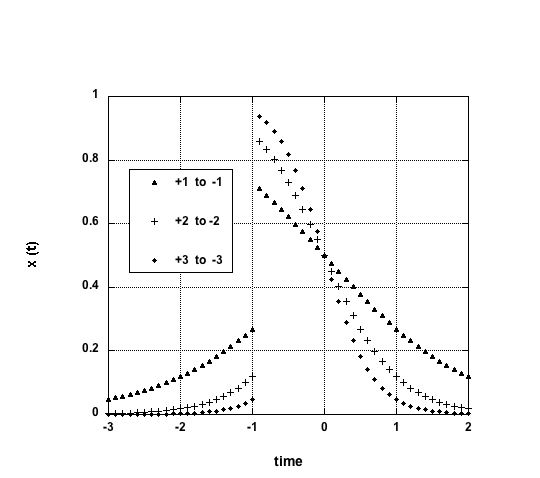}
\includegraphics[height=8cm,width=8cm]{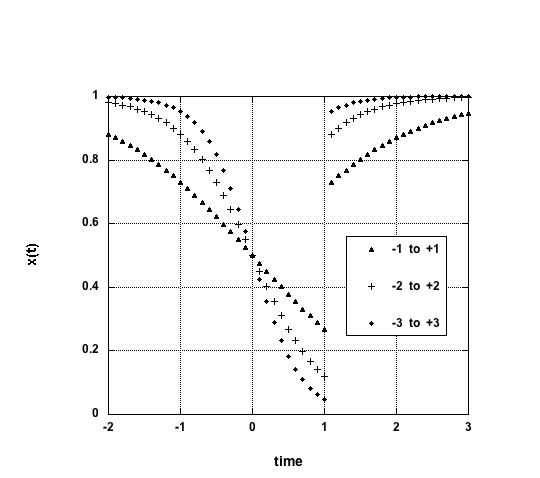}
\includegraphics[height=8cm,width=8cm]{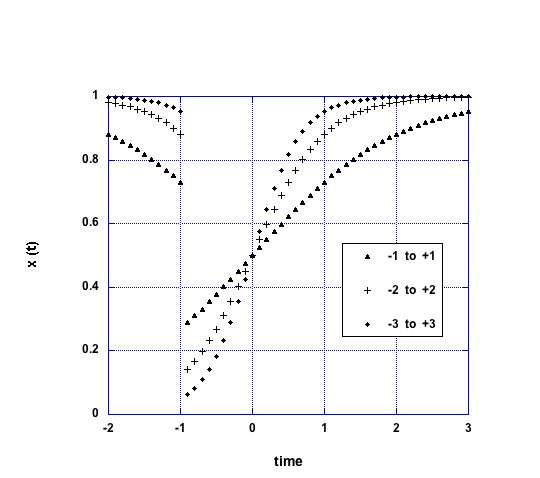}
\caption{\label{dropsjumpsV} The four possible cases of  population drops or jumps  in a  Verhulst  approach due to a drastic shift in $r$ values, see insert,   at some fixed  time $t$  (=+1 , or -1)}
\end{figure}

\begin{figure}
\includegraphics[height=8cm,width=8cm]{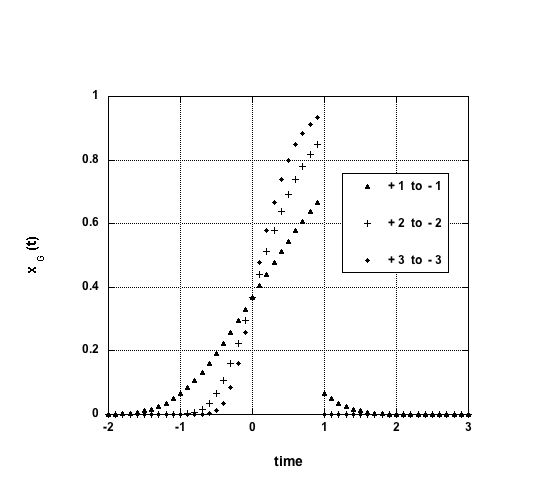}
\includegraphics[height=8cm,width=8cm]{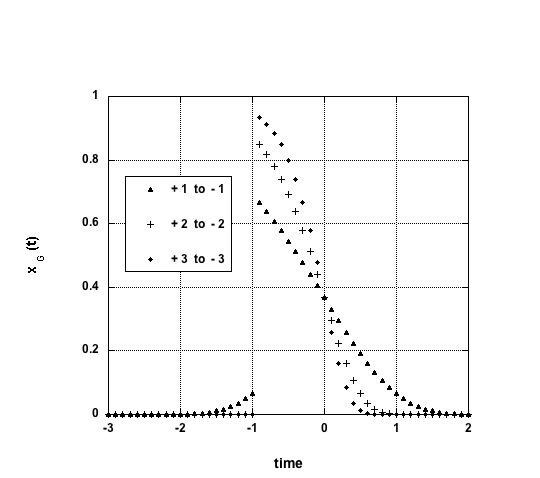}
\includegraphics[height=8cm,width=8cm]{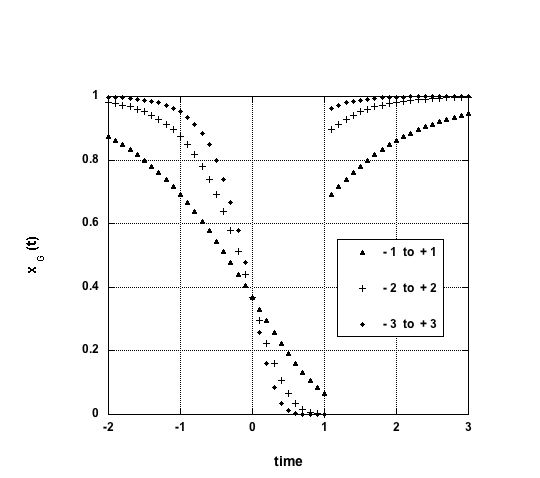}
\includegraphics[height=8cm,width=8cm]{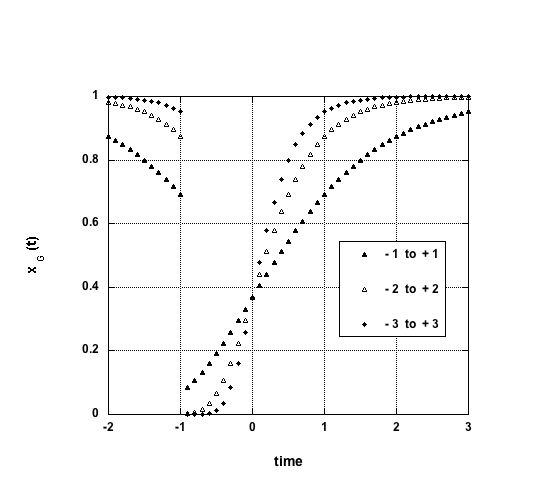}
\caption{\label{dropsjumpsG}The four possible cases of  population drops or jumps  in a Gompertz approach due to a drastic shift in $r$ values, see insert,   at some fixed  time $t$  (=+1 , or -1) }
\end{figure}

On the other hand, the four possible cases of  population drops or jumps  in a Verhulst approach due to a drastic shift in $r$ values, at some fixed  time $t$  are shown in Fig. \ref{dropsjumpsV}, as if a strong external event had influenced   a shift from some $r$ to some $-r$, or the contrary. Similar curves are shown for the Gompertz evolution law on Fig. \ref{dropsjumpsG}. 

\subsection{Time and size roles}
However such drastic change of signs of the growth rate are rather unlikely or very rough approximation of reality. A smoother evolution of $r$ must be searched for. In this respect,
let us recall that the dimension-prone Verhulst differential equation, Eq.(\ref{Verhulsteq19}),  is easily integrated,  
\begin{itemize}
\item if
the
growth rate $r$ is constant,  
\item  and if the
carrying capacity  $M $ is a constant.
\end{itemize}

Let us first observe that there are two equivalent ways to write both Verhulst and Gompertz equations, either emphasising the $time$ role or the $size$ role.

\subsubsection{Time role}
It is easy to observe that, after an appropriate simple change of variables, the corresponding differential equation to Eq.(\ref{Verhulst1diff}),  e.g.  for the skewed logistic, Eq.(\ref{skewedlog1integr}),  could read 
\begin{eqnarray} \label{skewedlog1}
\frac{dy}{dt}=\frac{b}{k}  \; y \;  \left[ k -  y \right]\; v(t)\;          ,
\end{eqnarray}
- thereby defining $v(t)$. One can easily integrate  both sides  of the equation, by the technique of separation of variables, i.e., 
\begin{eqnarray} \label{skewedlog11}
\frac{dy}{y \;  \left[ k -  y \right]}=  \frac{b}{k}  \; v(t) \;  dt    \;   ,     
\end{eqnarray}
and easily obtain the skewed logistic therefore.


\subsubsection{Size role}

Another interesting way, for further discussion, is to rewrite the evolution equation, letting the time $t$ in the right hand side be an implicit variable, thus referring to the evolution through the size itself, i. e. let \begin{eqnarray} \label{Vy}
\frac{dy}{dt}= V(y)   ,
\end{eqnarray}
where $V(y)= ry (1-y)$ in Verhulst model.
Imposing the stability condition $ \frac{dV(y)}{dt}= V(y)  =0   $,  one easily finds out that the r.h.s. is like \begin{equation} 
V(y) = a\;y + b\;y^2 + c\;y^3 +\;  ...., \\
\end{equation}
i.e. there is no zero order term.
In other words,  keeping the cubic term only, leads to
\begin{eqnarray} 
\frac{dy}{dt}\; = \;  ay (y-y_b)(y-y_c)\;,
\end{eqnarray}
therefore  simply generalising the Verhulst equation leading to the logistic map,  thereby introducing two inflexion points at $y_b$ and $y_c$. 

\subsection{Second order derivative}
 
 Without going into elementary considerations for mathematicians and physicists about the conditions on a differential equation to produce,  as its solution, a curve with an extremum, it is immediately obvious, in the present framework  that to differentiate once more   Eq.(\ref{Verhulst2sol})  and  Eq.(\ref{Gompertz2sol})   will  produce a maximum,  because of the introduction of a second inflexion point; - see Fig. \ref{V1G1} (rhs). However, the goal is to find  a  $first$ order differential equation (ODE)  not a $second$ ODE. As hinted here above, a time or size dependence of $r$ and/or $k$ ($ \sim M$) seems the simplest way.
 

\section{Growth rate role }

\subsection{Growth rate role in Verhulst model}



One may rewrite Verhulst equation, Eq.(\ref{Verhulst1diff}), explicitly expressing the $time$ dependence on the right hand side, i.e.
\begin{eqnarray}  \label{eq11}
 \frac{dx}{dt}=  \frac{re^{-rt}}{(1+e^{-rt})^2}\; =  \frac{r}{ 2(1+ cosh(rt))}\;.
\end{eqnarray}

In order to have a curve with an extremum the most simple way seems to consider its derivative; the extremum being located at the inflexion point of the logistic map. One finds
\begin{equation}\label{d2V}
\frac{d^2x}{dt^2}=r  \;  \left[ 1 -  2x \right]\; \frac{dx}{dt}\; = r^2 \; x\;(1-x)\;(1-2x),
\end{equation}

or emphasizing  the time dependence on the right hand side, i.e.   explicitly writing
\begin{equation}\label{f2}
\frac{d^2x}{dt^2}= \frac{r^2\; e^{rt}\;(1-e^{rt})}{(1+e^{rt})^3}\;.
\end{equation}

One can also rewrite the above as
\begin{equation}\label{Rx}
\frac{d^2x}{dt^2}= 
\left[ r^2 \frac{1-e^{rt}}{1+e^{rt}}\right]\; x\; (1-x) \; \equiv R_V \; x\; (1-x) 
\end{equation}

in order to reproduce the r.h.s. of Verhulst equation, but with a new $r$, called $R_V$ having a specific  time dependence, i.e. as
\begin{equation}\label{r2}
R_V 
=\left[ r^2 \frac{1-e^{rt}}{1+e^{rt}}\right] =   - r^2 \; tanh (rt)\; ,
\end{equation}
- which is not a birth-death growth rate of any known population, up to now.
In so doing one obtains the "generalized" time-Verhulst equation, admitting that the growth rate should be time dependent.



 

Considering a writing in terms of a size dependent $r$, one can write
\begin{eqnarray}\label{RxbV2}
\frac{d^2x}{dt^2}\equiv \frac{d}{dt}(\frac{dx}{dt})=
\rho_V\;   (\frac{dx}{dt})\; \; [1- (\frac{dx}{dt})]
\end{eqnarray}
 which has the Verhulst form $\frac{du}{dt}= \rho_V \; u \;  \left[ 1 -  u \right]$, with, comparing Eq.(\ref{d2V}) and Eq.(\ref{Rx}), 
\begin{eqnarray}\label{RxbV2rv}
\rho_V\;   =  \frac{R_V/r}{1-r\;x\; (1-x)}\;=\; \frac{r\;(1-2x)}{1-rx(1-x)}
\end{eqnarray}
where $\rho_V$ is expressed in terms of the population $size$. If one wishes  $\rho_V$  to be expressed in terms of the time $t$, one should replace $u$ ($\equiv \; x$)  and $R_V$ by  their expression found in Eq.(\ref{Verhulst2sol}) and Eq.(\ref{r2}) respectively; e. g., $R_V= r^2\; (1-2x(t))$.

\subsection{Growth rate role in Gompertz model}

Similar manipulations and algebra can be performed for the basic Gompertz model, i.e.,
\begin{equation}\label{RxbG}
\frac{d^2x}{dt^2} = r^2  
\left[   log\left[  \frac{k}{x} \right]-1\       \right]\;  x \;  log\left[  \frac{k}{x} \right]\; \equiv R_G \; x\;   log\left[  \frac{k}{x} \right]\;,
\end{equation}
thereby defining $R_G$, in terms of its size dependence,
  \begin{equation}\label{RxbGr}
  R_G= r^2  \left[   log\left[  \frac{k}{x} \right]-1\       \right]\; .
 \end{equation}

Therefore, in order to write
\begin{eqnarray}\label{RxbG2}
\frac{d^2x}{dt^2}\equiv \frac{d}{dt}(\frac{dx}{dt})=
\rho_G\;    (\frac{dx}{dt})\;     log\left[  \frac{k}{ (dx/dt)} \right]  
\end{eqnarray}
 which has the Gompertz form, 
 one has to introduce $\rho_G $
 with 
 \begin{eqnarray}\label{RxbG2rv}
\rho_G\;   =  \frac{R_G/r}{  log\left[k/x\right]  -log\left[r\;log(k/x)\right]}\;=r
\; \frac{\;log\left[k/x\right]-1 }{log\left[k/x\right]  -log\left[r\;log(k/x)\right]}
\end{eqnarray}
where $\rho_G$ is expressed in terms of the population size. If one wishes  $\rho_G$  to be expressed in terms of the time $t$, one should replace $R_G$  and $(dx/dt)$ by their expression given in Eq.(\ref{RxbGr}) and 
Eq.(\ref{Gompertz2sol}). It seems that such a writing is out of necessity here.

\section{Carrying capacity role}

Beside the possibility of expressing the growth rate in terms of some time or size dependence, let us recall that $k$  and $M$  in Gompertz and Verhulst model respectively are considered to be constant. To remove such a restriction has already been considered in works referenced to here above. However such a consideration is of highly relevant modern interest in our growing world, and thereby requests some attention.

\begin{figure}
\includegraphics[height=8cm,width=8cm]{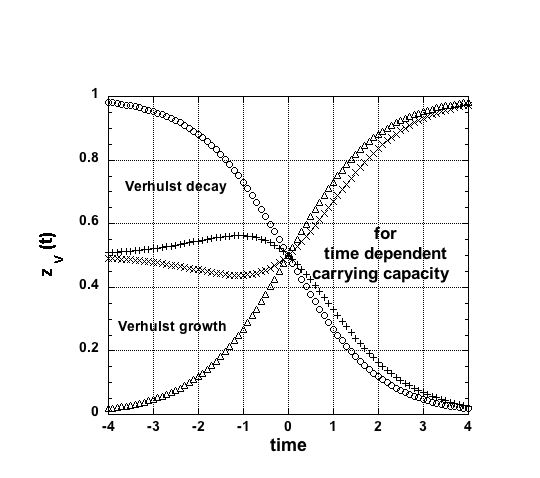}
\includegraphics[height=8cm,width=8cm]{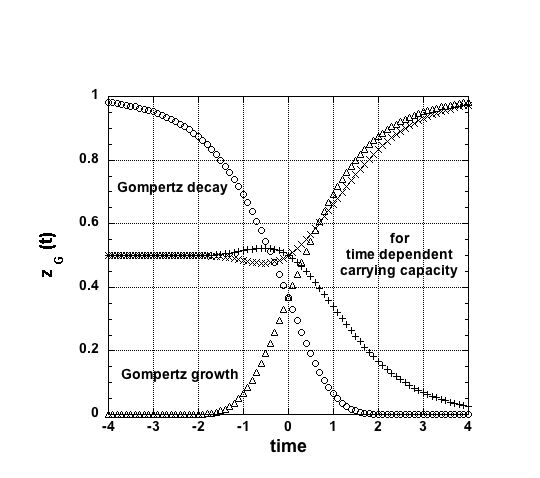}
\caption{\label{44} Possible cases of  population evolution  in a  (lhs) Verhulst  or a (rhs) Gompertz approach  when the carrying capacity has a   $time$ dependent component, as compared to an ordinary growth or decay evolution for constant $r  $ or $k$ ($\sim M$); $r=1$ for all  curves;   used analytic forms are given in the main text. A bump or a dip can be seen to occur here at finite $t \le0$ }
\end{figure}

Notice that some  population growing in a limited environment can approach the ultimate carrying capacity of that environment in several possible ways: (i)
it can adjust smoothly to an equilibrium, (ii) it can overshoot the limit and then dies back again  either in  a smooth  very damped way 
or  (iii) in an oscillatory, more or less damped,  way. Let it be emphasised that  the population
 can overshoot the carrying capacity and in the process decreases the ultimate carrying capacity by consuming some necessary nonrenewable resource:
e.g.,  deer or goats, when natural enemies are absent, often overgraze their range and cause erosion or destruction of the vegetation \cite{Kormandy69}

\subsection{World model}

Related considerations on the carrying capacity are  used in 
Meadows et al. \cite{Meadows} on
the "standard" world model. At first it assumes no major change in the physical, economic, or social relationships that have historically governed the development of the world system.
Food, industrial output, and population grow exponentially, as according to Malthus   \cite{Malthus798},   until the rapidly diminishing resource base forces a slowdown in industrial growth.
Population growth is finally halted by a rise in the death rate due to decreased food and medical services.  However, because of natural delays in the system,  the population (and pollution)
continues to increase for some time after the peak of industrialisation. 

Developments  of the world model have considered the doubling or resources  \cite{Meadows},  - together with plenty of other  interfering causes,  for describing  the population evolution.

 \subsection{Carrying capacity role in Verhulst and in Gompertz model}

 The time and/or size dependence of the so called carrying capacity  could be considered in the same spirit as the growth rate. However this is less trivial  and  apparently can only be tackled through numerical work. Indeed,  one can only easily separate the variables when one has either a form $k(x)$, or $k(t)$, in Gompertz model, or $M(N)$, or $M(t)$, in Verhulst model, in Eq.(\ref{Gompertz1dif}) and Eq.(\ref{skewedlog1}) respectively, i.e. 
 
 \begin{equation} \label {Gompertz1dif2}
\frac{dx}{dt}= r \; x(t) \;  log\left[  \frac{k(x(t))}{x(t)} \right]\;\\
\end{equation}
 \begin{equation} \label{skewedlog2}
\frac{dN}{dt}=\frac{r}{M(N(t))}  \; N(t) \;  \left[ M(N(t)) -  N(t) \right]\;     ,
\end{equation}
However to go on,  some analytic form for $k$ and $M$ is needed.
 This is leading to more elaborate, less universal,  algebraic work, and is outside our present report.

Nevertheless in order to substantiate the role of the carrying capacity  time dependence effect in the Verhulst and in Gompertz model, {\it a posteriori} analytic cases have been considered.
 A few results are shown in Fig. \ref{44} when a $time$ dependent behaviour is  {\it a posteriori} included as a mere factor into the original Verhulst of Gompertz laws, $r$ being kept constant.  For completeness, the used equations  in Fig. \ref{44} read
 \begin{eqnarray}\label{VGtanhCC}
z_V(t)=\frac{1}{2}\; [\frac{e^{ t}}{1+e^{ t}}]\;[\pm\frac{e^t-1}{e^t+1} +1]\\
z_G(t)= \frac{1}{2}[ e^{-e^{-t}} ] \;  [\pm \frac{ e^t-1}{ e^t+1} +1]\;.
\end{eqnarray}

 Bumps or dips are markedly seen,  within overall growth or decay features, even in this simple manipulation.
 A size dependent effect of the carrying capacity either in the Verhulst or Gompertz approach has not  been studied.
 
\section{Conclusion}



A goal has been to describe  some simple way toward one single simple  first order differential dynamics equation, recalling that 3-parameter families of curves often describe growth data adequately, like in Verhulst or Gompertz approach.
It has been stressed that  extensions of the corresponding equations  in order to introduce extrema, or more than one  inflexion point,  are easily found in the second order derivative. However a change of variables is necessary, implying that the growth rate or/and the carrying capacity becomes time (or size) dependent. This corresponds in practice at considering exogenous or/and endogenous causes, in other words,
external or internal field interactions with the population, i.e. intra-community interactions in the latter case.

 
Bass model \cite{Bass} and  Zwanzig model \cite{ZwanzigPNAS70} already contain such extrema in an {\it ad hoc} way within Verhulst  population growth framework.   Here it has been aimed at  providing  a single diffusion   Riccati equation with time/size dependent coefficients of realistic nature, enforcing that the corresponding first ODE is of the Gompertz or Verhulst type.  An analytic form for the growth rate coefficients has been derived. The carrying capacity form seems not so immediately tractable and demands more  numerical work.


Other ways of interest to describe a population evolution are those taking into account some time delay, as in the world model, and in work on  anticipation by Dubois \cite{Du1} or Rosen \cite{Ro1}. Indeed, in biological  and sociological terms, the logistic and Gompertz equations present a logical absurdity, since it requires that cause and effect occur $quasi$ simultaneously.


{\it In fine},  let it be pointed out that considerations of $discrete$ time effects do not seem to have been studied on Gompertz model, in contrast to the Verhulst case \cite{DubV200}: that might be of interest in the framework of CASYS  conference reports on chaotic, anticipatory or not,  systems.



 \noindent
 \textbf{\\Acknowledgements  }

Thanks go to Daniel. M. Dubois for inviting me to CASYS 2011, in some sense challenging me to write down coherently considerations which were not anticipated, but hidden for  long time in likely a few loosely connected neurones, ... and, as I discovered on the way,  in apparently old if not obscure references, for a physicist, sometimes.  Some support  by the COST Action MP0801 has allowed fruitful interactions, on the matter presented here above, with numerous colleagues; thanks to them as well.

 \noindent   \hskip 12pt
 
 {\bf Appendix A. Multiparameter extensions }
 
For completeness, it seems appropriate to recall generalisations of Verhulst and Gompertz laws, beside the so called skewed laws, mentioned in the main text.  E.g., one can write indeed
\begin{eqnarray} \label{G1V1}
G_1(t)= e^{-\rho \; e^{-t/\sigma}}       \\
V_1(t)= (1 + \rho\; e^{-t/\sigma})^{-\theta} \;,
\end{eqnarray}
where $\rho$, $\sigma$, and $\theta$ are to be positive numbers, - the logistic map corresponding to $\theta$=1.

Ahuja and Nash \cite{AhujaNash67} have  further generalised these distributions by introducing an additional parameter $ \phi> 0$. The density functions of these 
distributions are respectively  
 \begin{eqnarray} \label{AhujaNash67}
G_2(t;\rho,  \sigma, \phi ) =  \left[\frac{1}{\sigma\Gamma(\phi)}
\left(\rho \; e^{-t/\sigma}    \right)^{\phi}
\right] \; e^{-\rho \; e^{-t/\sigma}}       \\
V_2(t; \rho,  \sigma, \theta, \phi )= \left[ \frac{1}{\sigma B(\phi,\theta)}
\left(\rho \; e^{-t/\sigma}    \right)^{\phi}  \right] \; (1 + \rho\; e^{-t/\sigma})^{-(\theta+\phi)}.
\end{eqnarray}
However, it seems that  there is no physical justification of these parameters to be usefully discussed within the  present context.


\begin{thebibliography}{2}




\bibitem{Verhulst845}
P.F. Verhulst (1845)
 {\it  Nouveaux M\'emoires de l'Acad\'emie Royale des Sciences et Belles-Lettres  de Bruxelles} {\bf 18},  pp. 1-38. 

\bibitem{Malthus798} Th. R. Malthus  
(1798)  An Essay  on the Principle of Population as  It Affects the Future Improvement of Society.



\bibitem{Verhulst847}
P. F. Verhulst (1847)   
 {\it  Nouveaux m\' emoires de l'Acad\'emie Royale des Sciences et Belles
Lettres de Bruxelles} {\bf  20}, pp. 1-32.

  \bibitem{Avrami39}  M. Avrami (1939)  {\it J. Chem. Phys.} {\bf   7},  pp. 1103- 1112
  
  \bibitem{Lotka25}  A.J. Lotka (1925) Elements of Physical Biology. Williams \& Wilkins.

 \bibitem{Volterra31} V. Volterra (1931)   Le\c{c}on sur la Th\'eorie Math\' ematique de la Lutte pour la Vie. Gauthier-Villars.
 
 \bibitem{Blanco1} J.M.  Blanco (1993) 
 {\it Ecol. Model.} {\bf  66},  pp.  301-303.

  \bibitem{Gompertz825} R. Gompertz  (1825)  {\it Philos.  Trans. R.  Soc. London} {\bf  115},  pp. 513-585.

   \bibitem{PearlReed23}   R. Pearl and L. J. Reed (1925) 
 {\it Proc. Natl. Acad. Sci. USA} {\bf  11}, pp. 16-22.
 
\bibitem{OlshanskyCarnes}  
  S. J. 
  Olshansky and 
  B. A. Carnes (1997) 
  {\it  Demography} {\bf  34}, 
 pp. 1-15 
 
\bibitem{Gilipinetla66}  M.E. Gilpin, T.J. Case, and F J., Ayala (1976) 
 {\it  Math. Biosci.} {\bf 32}, pp. 131-140.

\bibitem{MontrollBadger74}  E.W. Montroll and W.W.  Badger (1974) Introduction  to Quantitative Aspects of Social Phenomena. Gordon and Breach.


 
  
   
   
    
 \bibitem{LeeWerning08} 
 A. H. Lee  and S.
  Werning (2008)    {\it Proc. Natl. Acad. Sci. USA} {\bf 105}, pp. 582-587.
 
 \bibitem{vonBertalanffy}  L. von Bertalanffy (1938) 
 {\it Human Biol. } {\bf 10}, pp. 181-213.

\bibitem{Pearletal28f} 
 R. Pearl,  C. P. Winsor, and F. B. White  (1928)
 {\it Proc. Natl. Acad. Sci. USA} {\bf 14}, pp. 895-901, 

\bibitem{Pearletal28a} R. Pearl,  A. A. Winsor, and J. R. Miner (1928) 
 {\it  Proc. Natl. Acad. Sci. USA} {\bf  14}, pp. 1-4.

\bibitem{auslooscrystalgrowth} R. Cloots, N. Vandewalle, and M. Ausloos (1996) 
{\it J. Cryst. Growth} {\bf 166}, pp.   816-819.

 \bibitem{DeVriesetal03} G. de Vries, T. Hillen, M. Lewis, M. Li, J. M\"uller, and B. Sch\"onfisch (2003) A Short Course in Mathematical and Computational
Biology. SIAM Publishing 
 
  \bibitem{hayw99} J. Hayward (1999)  
   {\it  J. Math. Sociol.} {\bf 23}, pp.  255-292. 
  
\bibitem{hayw05}  J. Hayward (2005) 
{\it  J. Math. Sociol.} {\bf 29}, pp.  177-207.
 

  \bibitem{religion1} M. Ausloos and  F. Petroni (2007)  
  {\it Europhys. Lett. } {\bf 77}, pp. 38002 (4pp) 
  
   \bibitem{religion2} M. Ausloos and  F. Petroni (2009)
    {\it Physica A} {\bf   388}, pp.  4438-4444. 
    
    \bibitem{religion566} M. Ausloos and F. Petroni (2010) On World Religion Adherence Distribution Evolution, 
    in  Econophysics Approaches to Large-Scale Business Data and Financial Crisis, Editors: M. Takayasu, T. Wanatabe, H. Takayasu. Springer, pp. 289-312.  
 
\bibitem{religion568} M. Ausloos (2010) On religion and language evolutions seen through mathematical and agent based models, in Proceedings of the First Interdisciplinary CHESS Interactions Conference,  C. Rangacharyulu and E. Haven, Eds.  World Scientific,  pp. 157-182 

     \bibitem{NowakKrakauerPNAS99} M. Nowak and D. Krakauer (1999) 
  {\it Proc. Natl. Acad. Sci. USA} {\bf  96}, pp.  8028-8033.

\bibitem{Abr+03} D.M. Abrams and S.H. Strogatz  (2003) 
  {\it Nature} {\bf 424},  p. 900.
  
  \bibitem{Makeham} W. M. Makeham (1860)  {\it  J. Inst. Actuaries and Assur. Mag.} {\bf  6}, pp. 301-310. 
  
  
  
  
 \bibitem{Trachtenberg} H.L. Trachtenberg (1924) 
 {\it  J. Royal Stat. Soc.} {\bf  87},
 pp. 278-290.

 
 \bibitem{Beeton70}  A. M. Beeton (1970) Statement on Pollution and Eutrophication of the Great Lakes, The University of Wisconsin Center for Great Lakes Studies Special Report \#11 
 Univ.  Wisconsin.

\bibitem{Meadows} D.H. Meadows,  D.L.  Meadows, J.  Randers, and  W. W. Behrens III (1972) The limits to Growth, A Report for THE CLUB OF ROME'S Project on the Predicament of Mankind. Universe Books.

\bibitem{Kormandy69} E.J. Kormandy   (1969) Concepts of Ecology. 
Prentice-Hall,  pp. 95-97.

\bibitem{ormerodPhA293} P. Ormerod and C. Mounfield  (2001)
 {\it Physica A} {\bf 293}, pp. 573-582.

\bibitem{miskiewiczauslooscycles}  J. Miskiewicz, and M. Ausloos  (2004) 
 {\it Physica A} {\bf 336}, pp. 206-214. 

\bibitem{sanglierauslooscycles} M. Ausloos, J. Miskiewicz, and M. Sanglier (2004)
 {\it Physica A} {\bf  339},  pp. 548-558.

\bibitem{Wright05} I. Wright (2005)
 {\it Physica A} {\bf  345}, pp. 608-610.

\bibitem{redelicoausloos08} F. O. Redelico, A.N. Proto, and M. Ausloos (2008) 
 {\it Physica A} {\bf 387},  pp. 6330-6336 


  \bibitem{AhujaNash67} J. C. Ahuja and  S. W. Nash (1967) 
{\it  Sankhya A} {\bf 29}, pp. 141-156. 
 
\bibitem{Bass}   F.M. Bass (1969)  
 {\it Management Science} {\bf  18}, pp. 215-227.

 \bibitem{ZwanzigPNAS70}
 R. Zwanzig  (1973) 
 {\it Proc. Natl. Acad. Sci. USA} {\bf 70}, pp.  3048-3051.
 
 \bibitem{MontrollPNAS69}  E. W. Montroll  (1972) 
 {\it Proc. Natl. Acad. Sci. USA} {\bf 69}, pp. 2532-2536.
 
 
 \bibitem {Du1} D.M. Dubois (2000) Review of Incursive, Hyperincursive
and Anticipatory Systems - Foundation of Anticipation in Electromagnetism. Computing Anticipatory Systems, in  CASYS'99, D. M. Dubois, Ed.,  The American Institute of Physics, AIP Conference Proceedings {\bf 517}, pp. 3-30. 

\bibitem {Ro1} R. Rosen (1985) Anticipatory Systems. Pergamon Press.

\bibitem{DubV200} D. Dubois (2006) Recurrent generation of Verhulst chaos maps at any order and their stabilization diagram by anticipative control,  in The Logistic Map and the Route to Chaos,  M. Ausloos  \& M. Dirickx,  Eds., Springer-Verlag, Berlin/Heidelberg, pp. 53Ð75.

\end{thebibliography}
\end{document}